\documentclass[aps,prl,twocolumn,superscriptaddress,showpacs,preprintnumbers,amsmath,amssymb]{revtex4}

\usepackage{graphicx}% Include figure files
\usepackage{dcolumn}% Align table columns on decimal point
\usepackage{bm}% bold math
\usepackage{pslatex}
\newcommand{\moeller}{M\o ller}
\newcommand{\cherenkov}{Cerenkov}
\newcommand{\targetreynoldsnumber}{2 \times 10^{5}}
\newcommand{\solidangle}{0.62}      % solid angle of the pbf2 calorimeter,
\newcommand{\xymodriftspace}{7.21} % length in meters of te drift space between the two xymos

\newcommand{\experimentalasymmetryalucor}{-5.44}
\newcommand{\statisticalerroralucor}{0.54}   % experimental
\newcommand{\polarisationmeasurementerror}{0.11}  % ppm entspricht 2% experimental
\newcommand{\polarisationinterpolationerror}{0.19}  % ppm experimental
\newcommand{\polarisationmeasurementcorrection}{-1.07}  % ppm experimental
\newcommand{\polarisationinterpolationcorrection}{0.00}  % experimental
\newcommand{\combinedsyspolerroralucor}{0.26}  % experimental
\newcommand{\aluminumwindowthickness}{250} % thickness in mu m
\newcommand{\aluminumwindowcorectionfactor}{1.030} % divide measured asymmetry by this factor to get phys asymmetry
\newcommand{\aluminumwindowcorectionfactorerror}{0.003} % divide measured asymmetry by this factor
\newcommand{\aluminumwindowcorrection}{0.16}      % absolute correction in ppm
\newcommand{\aluminumwindowcorrectionerror}{0.02}      % absolute correction in ppm
\newcommand{\qsquaredavaraged}{0.230}  % including rad cor four momentum transfer averaged over detector acceptance
\newcommand{\qsquaredhappex}{0.477}  % four momentum transfer averaged over detector acceptance
\newcommand{\standardmodelasymmetryaveraged}{-6.30}         % averaged with cross section over pbf2 detector incl em rad cor
\newcommand{\standardmodelasymmetrytotalerror}{0.43} %ppm
\newcommand{\gesplusgmsaveragedfactor}{0.225}   % incl alu corr incl rad corr GES + \gesplusgmsfactor GMS = \gesplusgms +- gesplusgmserror, avereaged over pbf2 detector
\newcommand{\gesplusgmsaveraged}{0.039}         % incl alu corr GES + \gesplusgmsfactor GMS = \gesplusgms +- gesplusgmserror, avereaged over pbf2 detector
\newcommand{\gesplusgmsaveragederror}{0.034}    % GES + \gesplusgmsfactor GMS = \gesplusgms +- gesplusgmserror, avereaged over pbf2 detector

\newcommand{\happexangle}{12.3}
\newcommand{\gesplusgmserrorhappex}{0.026}    % newly calculated GES + \gesplusgmsfactor GMS = \gesplusgms +- gesplusgmserror, avereaged over pbf2 detector
\newcommand{\gesplusgmsfactorhappex}{0.395}   % GES + \gesplusgmsfactor GMS = \gesplusgms +- gesplusgmserror, avereaged over pbf2 detector
\newcommand{\gesplusgmshappex}{0.034}         % GES + \gesplusgmsfactor GMS = \gesplusgms +- gesplusgmserror, avereaged over pbf2 detector
\newcommand{\fonesplusftwoserrorhappex}{0.019}    % GES + \gesplusgmsfactor GMS = \gesplusgms +- gesplusgmserror, avereaged over pbf2 detector
\newcommand{\fonesplusftwosfactorhappex}{0.186}   % GES + \gesplusgmsfactor GMS = \gesplusgms +- gesplusgmserror, avereaged over pbf2 detector
\newcommand{\fonesplusftwoshappex}{0.024}         % GES + \gesplusgmsfactor GMS = \gesplusgms +- gesplusgmserror, avereaged over pbf2 detector
\newcommand{\fonesplusftwosaveragedfactor}{0.130}
\newcommand{\fonesplusftwosaveraged}{0.032}
\newcommand{\fonesplusftwosaveragederror}{0.028}
\newcommand{\gmsassumption}{-0.099}
\newcommand{\gmsassumptionerror}{0.154}
\newcommand{\ftwosassumption}{-0.150}
\newcommand{\ftwosassumptionerror}{0.150}
\newcommand{\finalgesafour}{0.061}
\newcommand{\finalgesafourerror}{0.035}

\newcommand{\finalgescombined}{0.066}
\newcommand{\finalgescombinederror}{0.026}
\newcommand{\finalgessigma}{2.5}
\newcommand{\finalfonesafour}{0.052}
\newcommand{\finalfonesafourerror}{0.034}

\newcommand{\finalfonescombined}{0.052}
\newcommand{\finalfonescombinederror}{0.024}
\newcommand{\finalfonessigma}{2.2}
\newcommand{\strangemassofthenucleon}{50}
\newcommand{\strangespinofthenucleon}{-10}
\newcommand{\radcorasymreduction}{1.3} % reduction af asymmetry (betrag wird kleiner), unit is \%
\newcommand{\sinthetaweffective}{0.23113(15)}   % effective sin2thetaW in MSbar
\newcommand{\MAMIenergy}{854.3}

\newcommand{\moellererror}{2.1}         % unit is \%
\newcommand{\moellerfinalpolaerror}{4}  % unit is \% absolute
\newcommand{\doublehitlosses}{1}        %
\newcommand{\lowerenergycut}{1.6}       % unit is sigma energy resolution
\newcommand{\upperenergycut}{2.0}       % inits is sigma energy resoltution
\newcommand{\gammafrompizerodilutionfactor}{1} % unit is \%, i.e. much less than 1\%
\newcommand{\gammafrompizerodilutioncorrection}{0.00} % unit is \%, i.e. much less than 1\%
\newcommand{\gammafrompizerodilutioncorrectionerror}{0.06} % error in units of 1\%
\newcommand{\totalnumberofhistograms}{7.3}
\newcommand{\totalnumberofevents}{4.8}
\newcommand{\budgetstatisticcorr}{}
\newcommand{\budgetstatisticerr}{\statisticalerroralucor}
\newcommand{\budgettargetdensLcorr}{0.58}
\newcommand{\budgettargetdensLerr}{0.09}
\newcommand{\budgettargetdensIcorr}{0.00}
\newcommand{\budgettargetdensIerr}{0.04}
\newcommand{\budgetLUMOnonlincorr}{0.30}
\newcommand{\budgetLUMOnonlinerr}{0.04}
\newcommand{\budgetquadcorr}{-0.11}
\newcommand{\budgetquaderr}{0.08}
\newcommand{\budgetasymIcorr}{0.64}
\newcommand{\budgetasymIerr}{0.04}
\newcommand{\budgetenergycorr}{-0.05}
\newcommand{\budgetenergyerr}{0.02}
\newcommand{\budgetposcorr}{-0.03}
\newcommand{\budgetposerr}{0.02}
\newcommand{\budgetangcorr}{0.03}
\newcommand{\budgetangerr}{0.03}

\begin{document}
% \preprint{}
\title{Measurement of Strange Quark Contributions to the Nucleon's Form Factors
at $Q^2$=\qsquaredavaraged~(GeV/c)$^2$}
\author{F.~E.~Maas}
\email[corresponding author: ]{ maas@kph.uni-mainz.de}
\author{P.~Achenbach}
\author{K.~Aulenbacher}
\author{S.~Baunack}
\author{L.~Capozza}
\author{J.~Diefenbach}
\author{K.~Grimm}
\author{Y.~Imai}
\author{T.~Hammel}
\author{D.~von Harrach}
\author{E.-M.~Kabu{\ss}}
\author{R.~Kothe}
\author{J.~H.~Lee}
\author{A.~Lorente}
\author{A.~Lopes Ginja}
\author{L.~Nungesser}
\author{E.~Schilling}
\author{G.~Stephan}
\author{C.~Weinrich}
\affiliation{Institut f\"ur Kernphysik,
             Johannes Gutenberg Universit{\"a}t Mainz,
             J. J. Becherweg 45,
             D-55099 Mainz,
             Germany}
\author{I.~Altarev}
\affiliation{St. Petersburg Institute of Nuclear Physics, Gatchina, Russia}
\author{J.~Arvieux}
\author{B.~Collin}
\author{R.~Frascaria}
\author{M.~Guidal}
\author{R.~Kunne}
\author{D.~Marchand}
\author{M.~Morlet}
\author{S.~Ong}
\author{J.~van de Wiele}
\affiliation{Institut de Physique Nucleaire,
             91406 - Orsay Cedex,
             France}
\author{S.~Kowalski}
\author{B.~Plaster}
\author{R.~Suleiman}
\author{S.~Taylor}
\affiliation{%Physics Department and
             Laboratory for Nuclear Science,
             Massachusetts Institute of Technology,
             Cambridge, MA 02139, USA}
\date{\today}
\begin{abstract}
We report on a measurement of the parity-violating asymmetry in the scattering
of longitudinally polarized electrons on unpolarized protons at a $Q^2$
of \qsquaredavaraged~(GeV/c)$^2$
and a scattering angle of $\theta_e = 30^{\circ} - 40^{\circ}$.
Using a large acceptance fast PbF$_2$ calorimeter
with a solid angle of $\Delta\Omega = \solidangle$~sr the A4 experiment is the first
parity violation experiment to count individual scattering events.
The measured asymmetry is
$A_{\rm phys} =(\experimentalasymmetryalucor \pm
                 \statisticalerroralucor_{\rm stat}\pm
                 \combinedsyspolerroralucor_{\rm sys}) \times 10^{-6}$.
The Standard Model expectation assuming no strangeness contributions to
the vector form factors is $A_0=(\standardmodelasymmetryaveraged \pm
                                        \standardmodelasymmetrytotalerror) \times 10^{-6}$.
The difference is a direct measurement of the strangeness contribution
to the vector form factors of the proton. The extracted value is
$G^s_E + \gesplusgmsaveragedfactor G^s_M = \gesplusgmsaveraged \pm \gesplusgmsaveragederror$ or
$F^s_1 + \fonesplusftwosaveragedfactor F^s_2 = \fonesplusftwosaveraged \pm \fonesplusftwosaveragederror$.
\end{abstract}
% \keywords{}
\maketitle
\begin{figure}%[htb]
    \begin{center}
           \includegraphics[width=0.45\textwidth]{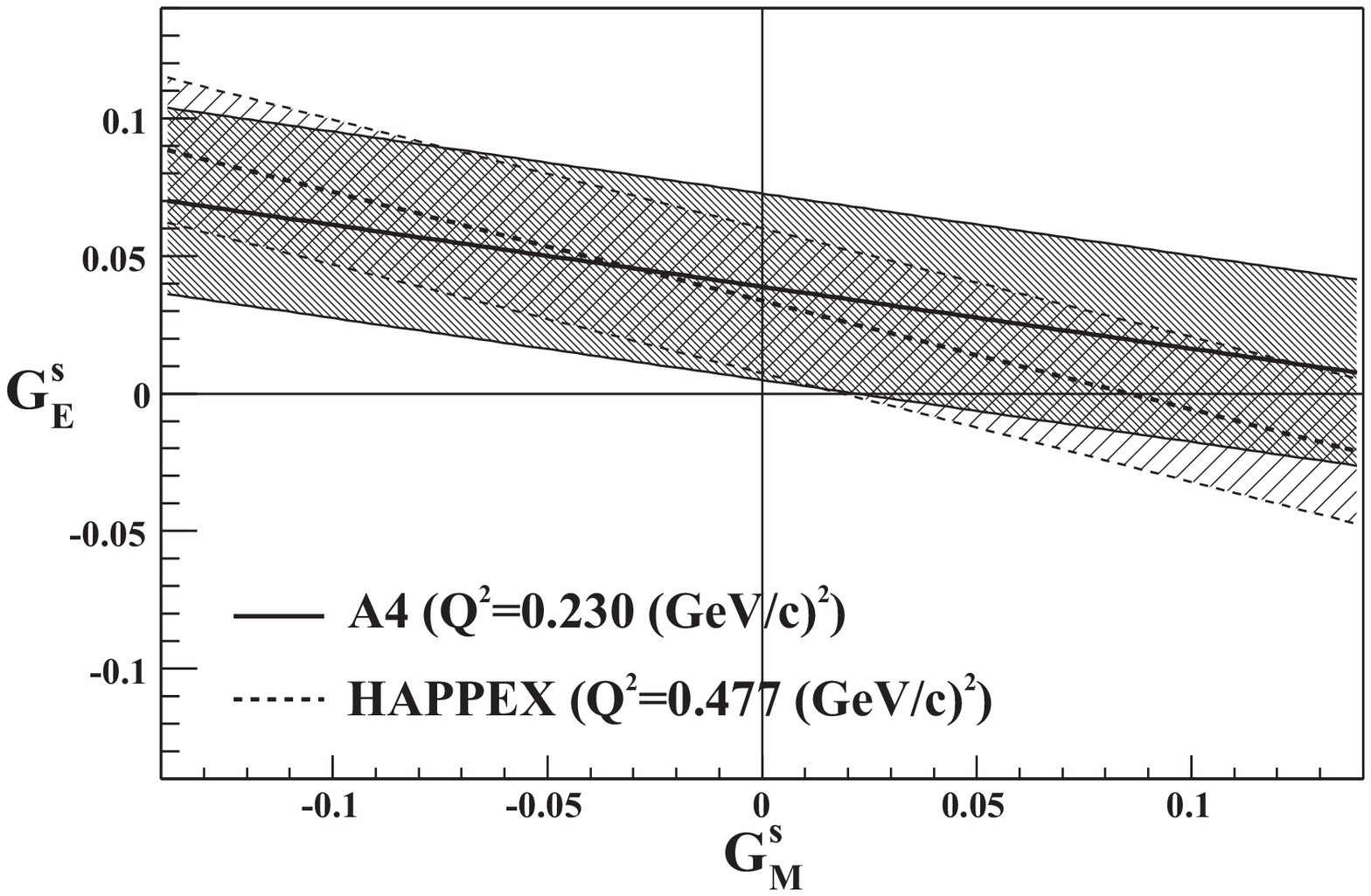}\\
      \caption{The solid line represents all possible combinations of $G^s_E + \gesplusgmsaveragedfactor G^s_M$
               as extracted from the work presented here at a $Q^2$ of $\qsquaredavaraged$~(GeV/c)$^2$.
               The densely hatched region represents the uncertainty. The recalculated result from the HAPPEX
               published asymmetry at $Q^2$ of $\qsquaredhappex$~(GeV/c)$^2$ is indicated by the dashed line,
               the less densely hatched area represents the associated error of the HAPPEX result.}\label{fig:gegmcombined}
    \end{center}
\end{figure}

%
%
% Motivation
%
%
The understanding of sea-quark degrees of freedom of the nucleon
and the interaction of the quarks with the vacuum in the nonperturbative
low energy regime of quantum chromo\-dynamics (QCD) is very poor even today.
Since the nucleon has no net strangeness,
any contribution of strange quarks to the nucleon structure observables
is of particular interest because it would be a pure sea-quark effect.
For example the scalar strangeness content of the nucleon that gives a contribution
to the mass of the nucleon has been discussed in the context of the $\Sigma$ commutator
which can be related to the $\pi$-N scattering amplitude  \cite{sigmaterm:olsson:02}.
New measurements suggest a large contribution
of strange quarks to the mass of the nucleon on the order of $\strangemassofthenucleon\,\%$.
The interpretation of the unexpected nucleon spin
content results  \cite{spinstructure:adams:97} suggests a sizeable contribution,
$\Delta s\approx \strangespinofthenucleon\,\%$, of the strange quarks to the nucleon
spin.\\
%
%
% Other Experiments
%
%
Estimates of the strange quark contribution to the magnetic
and electric vector form factors predict sizeable effects
accessible to experiments  \cite{strangereview:beck:01,
strangeness:silva:03}. Recently, two experiments,
SAMPLE at Bates  \cite{sample:hasty:00} and HAPPEX at TJNAF  \cite{happex:aniol:01},
have explored parity-violating (PV) asymmetries on the
proton and the deuteron in two different kinematical regions.
We report here on a new measurement at a four momentum transfer
$Q^2$ of $\qsquaredavaraged$~(GeV/c)$^2$ at the
Mainzer Mikrotron accelerator facility (MAMI) \cite{MAMI:euteneuer:94}.
%
%
% New and complementary experiment
%
%
The A4 experiment at MAMI is complementary to other experiments
for two reasons. Firstly its $Q^2$-value tests models predicting
an enhanced strangeness contribution  at this point \cite{strangeness:weigel:95}
and secondly, for the first time counting techniques
are used in a scattering experiment measuring
a PV asymmetry. Therefore possible systematic contributions to
the experimental asymmetries and the associated uncertainties are of a different
nature as compared to previous experiments, which use analogue integrating
techniques.\\
%
% formula for parity violation and strange quarks
%
Access to the strangeness nucleon vector current matrix elements is possible
in the framework of the standard model by a measurement of
the weak vector form factors $\tilde{G}^p_{E,M}$ of the proton \cite{pvstrangeness:kaplan:88}.
They can be expressed in terms of the known nucleon
electro-magnetic vector form factors $G_{E,M}^{p,n}$ and the unknown
strangeness contribution $G_{E,M}^s$.
%{ \sc The neutral current weak vector form factors of the proton $\tilde{G}^p_{E,M}$
%form together with the axial current of the lepton
%parity odd amplitudes in the elastic scattering of longitudinally polarized electrons on
%unpolarized protons.}
The interference between weak (Z$^0$) and electro-magnetic ($\gamma$)
amplitudes leads to a PV asymmetry $A_{LR}(\vec{e}p)$ in the elastic scattering cross section
for right- and left-handed electrons ($\sigma_R$ and $\sigma_L$ respectively),
which is given in the framework of the Standard Model \cite{pvsummary:musolf:94} and can be
expressed as a sum of three terms, $A_{LR}(\vec{e}p) = A_V + A_s + A_A $, with
\begin{eqnarray}
A_V              &=& - a \,
                      \rho_{eq}' \{(1 - 4 \hat{\kappa}_{eq}' \hat{s}^2_Z )
                   -
                   \frac{\epsilon G_E^p G_E^n + \tau G_M^p G_M^n}{\epsilon (G_E^p)^2 + \tau (G_M^p)^2}  \},
                 \label{eq:asymmtheory} \\
A_s              &=&  a \,
       \{ \rho_{eq}'
                        \frac{\epsilon G_E^p G_E^s + \tau G_M^p G_M^s}{\epsilon (G_E^p)^2 + \tau (G_M^p)^2} \},\\
A_A              &=&  a \,
   \{ \frac{(1-4\hat{s}^2_Z)\sqrt{1-\epsilon^2}\sqrt{\tau (1+\tau)} G_M^p \tilde{G}_A^p}
   {\epsilon (G_E^p)^2 + \tau (G_M^p)^2}\}.
\end{eqnarray}
%
% explanation of the formula
%
$A_{V}$ represents the vector coupling on the proton vertex where the possible
strangeness contribution has been taken out and has been put into $A_s$.
$A_s$ is a term arising only if a contribution from strangeness to the
electro-magnetic vector form factors is present and the
term $A_A$ represents the contribution from the axial coupling at the proton
vertex due to the neutral current weak axial form factor $\tilde{G}_A^p$.
The quantity $a$ represents $(G_{\mu} Q^2) /(4 \pi \alpha \sqrt{2})$.
$G_{\mu}$ is the Fermi coupling constant as derived from muon decay.
$\alpha$ is the fine structure constant,
$Q^2$ the negative square of the four momentum transfer,
$\tau = Q^2/(4 M_p^2)$ with $M_p$ the proton mass and
$\epsilon = [1 + 2 (1 + \tau)\tan^2(\theta_e/2)]^{-1}$
with $\theta_e$ the laboratory scattering angle of the electron.
The electro-magnetic form factors $G_{E,M}^{p,n}$ are taken from a recent
parametrization (version 1, page 5) by Friedrich and Walcher \cite{emformfactor:friedrich:03},
where we assign an experimental error of $3\,\%$ to $G_M^{p}$ and $G_E^p$,
of $5\,\%$ to $G_M^n$, and of $10\,\%$ to $G_E^n$.
%
% electro-weak radiative corrections
%
Electro-weak radiative corrections are included in the factors $\rho_{eq}'$
and $\hat{\kappa}_{eq}'$ which have been evaluated in the $\overline{MS}$
renormalization scheme \cite{ewcorrections:marciano:84}.
%In the one boson exchange approximation (tree level) the parameters
%$\rho_{eq}'$ and $\hat{\kappa}_{eq}'$ are equal to 1.
We use a value for $\hat{s}^2_Z = \sin^2 \hat{\theta}_W(M_Z)_{\overline{MS}}$ of
\sinthetaweffective\   \cite{pdg:hagiwara:02}.
The electro-weak radiative corrections to $A_A$ have also been estimated within
the $\overline{MS}$ renormalization scheme and
the radiative corrections given in  \cite{axialformfactor:zhu:00} as well
as a value of $\Delta s = -0.1\pm0.1$ are included
in $\tilde{G}_A$.
%
% Electro-magnetic radiative corrections (internal, external)
% and energy loss due to ionization
%
Electro-magnetic internal and external radiative corrections to the asymmetry
and the effect of energy loss due to ionization in the target have been
calculated. They reduce the expected asymmetry in our kinematics by about
$\radcorasymreduction\,\%$.
%{ \sc The largest contribution comes from the reduction of the
%electron energy at the interaction point.}
%
% Value for A_0
%
%{ \sc Including only the electro-weak radiative corrections, we obtain for comparison
%a Standard Model value for the asymmetry without strangeness contribution at
%the beam energy of $E_e=\MAMIenergy$~MeV and a $\theta_e = 35^{\circ}$ of
%$A_0' = A_{V} + A_{A} =
%  (\standardmodelasymmetrythirtyfivevector \pm \standardmodelasymmetryvectorerror)
%+ (\standardmodelasymmetrythirtyfiveaxial  \pm \standardmodelasymmetryaxialerror)$~ppm at a
%$Q^2 = \qsquarednotavaraged$~GeV$^2$). }
In order to compare to the measured asymmetry, we include the effects of electro-magnetic
radiative corrections and energy loss in the target and average $A_0 = A_{V} + A_{A}$
over the acceptance of the detector and the target length. We obtain
the expected value for the asymmetry at the averaged $Q^2$ without
strangeness contribution to the vector form factors of
$A_0(Q^2=\qsquaredavaraged{\rm (GeV/c)}^2) =
\standardmodelasymmetryaveraged \pm \standardmodelasymmetrytotalerror$~ppm.\\
%
% A4 Experiment
%
The PV asymmetry was measured at the MAMI accelerator facility in Mainz \cite{MAMI:euteneuer:94}
using the setup of the A4 experiment \cite{a4experi:maas:03}.
%{ \sc A 20~$\mu$A beam of $\MAMIenergy$~MeV longitudinally polarized electrons interacts with
%a 10~cm long high power liquid hydrogen target. The longitudinal electron polarization was
%typically about 80\%. Scattered electrons are detected by a fast, homogeneous,
%totally absorbing, segmented PbF$_2$ calorimeter with a solid angle
%of $\Delta \Omega = \solidangle$~sr.
%As shown in fig. \ref{fig:a4spektrum} the energy of accepted particles was measured
%and histogrammed separately for each submodule number and
%for the two electron beam helicities.
%In the energy histograms the elastic
%scattering events have been separated in the analysis from the inelastic channels to compute
%the raw PV asymmetry.}\\
%
%
%
% Source
%
The polarized $\MAMIenergy$~MeV electrons were produced using a strained layer
GaAs crystal which is illuminated with circularly polarized laser
light \cite{polsource:aulenbacher:97}. Average beam polarization was about $80\,\%$.
The helicity of the electron beam has been selected every 20.08~ms
by setting the high voltage of a fast Pockels cell according to a pattern of
four helicity states, either $(+P -P -P +P)$ or $(-P +P +P -P)$.
The pattern was selected randomly
by a pseudo-random bit generator. A 20~ms time window enabled the
histogramming in all detector channels and an integration circuit
in the beam monitoring and luminosity
monitoring systems. The exact window length was locked
to the power frequency of 50~Hz in the laboratory by a phase locked loop.
For normalization, the gate length was measured for each helicity.
Between each 20~ms measurement gate, there was an 80~$\mu$s time
window for the high voltage at the Pockels cell to be changed.
The intensity $I =$~20~$\mu$A of the electron current was stabilized
to better than $\delta I/I \approx 10^{-3}$. An additional $\lambda/2$-plate in the
optical system was used to rotate small remaining linear polarization components and
to control the helicity correlated asymmetry in the electron beam current
to the level of $< 10$~ppm in each five min run.\\
%
% Beam monitor and stab. system, beam properties
%
%{ \sc From the source to the target, the electron beam develops fluctuations
%in beam parameters such as position, energy and intensity which are
%partly correlated to the reversal of the helicity from +P to -P. }
We have used a system of microwave
resonators in order to monitor beam current, energy, and position in two
sets of monitors separated by a drift space of about $\xymodriftspace$~m in front of the
hydrogen target.
In addition, we have used a system of 10 feed-back loops in order to stabilize
current, energy  \cite{estabil:seidl:00}, position, and angle of the beam.
%
% Polarisation measurement (Moeller polarimeter)
%
The polarization of the electron beam was measured using a
\moeller\ polarimeter which is located on a beam line in
another experimental hall.
This \moeller\ polarimeter \cite{moellerpol:bartsch:01} has an accuracy of
$\moellererror\,\%$. Due to the fact, that we have to interpolate between
the weekly \moeller\ measurements, the uncertainty in the
knowledge of the beam polarization increased to
$\moellerfinalpolaerror\,\%$.
%
% hydrogen target and luminosity monitors
%
The 10~cm high power, high flow liquid hydrogen target was optimized to guarantee
a high degree of turbulence with a Reynolds-number of $R > \targetreynoldsnumber$
in the target cell in order to increase the effective
heat transfer. This new technique allowed us for the first time to avoid
a fast modulation of the beam position of
the intense cw 20~$\mu$A beam and it also allowed us to stabilize the beam position on the
target cell without target density fluctuations arising from boiling.
The total thickness of the entrance and exit aluminum windows is \aluminumwindowthickness~$\mu$m.
The luminosity $L$ was monitored for each helicity state (R, L) during the experiment using eight
water-\cherenkov\ detectors (LuMo) that detect scattered particles symmetrically around the
electron beam for small scattering
angles in the range of $\theta_e = 4^{\circ}-10^{\circ}$, where  the PV asymmetry is
negligible. The photomultiplier tube currents of these luminosity detectors are
integrated during the 20~ms measurement period by gated integrators and then digitized by customized 16-bit
analogue-to-digital converters (ADC). The same method was used for all the beam parameter signals
and these data were stored with the helicity information as a function of time
for each 20~ms time window.
A correction was applied for the nonlinearity of the
luminosity monitor photomultiplier tubes. This was measured and verified separately
by varying the beam current from 0-23~$\mu$A several times per week.
From the beam current helicity pair data $I^{R, L}$ and luminosity monitor helicity pair
$L^{R, L}$ data we calculated
the target density $\rho^{R, L} = L^{R, L}/I^{R, L}$ for the two helicity states
independently for the analysis of the data.\\
%
%  detector and electronics
%
As particle detector we developed a new type of a very fast, homogeneous, total absorption
\begin{figure}%[htb]
    \begin{center}
      \includegraphics[width=0.45\textwidth]{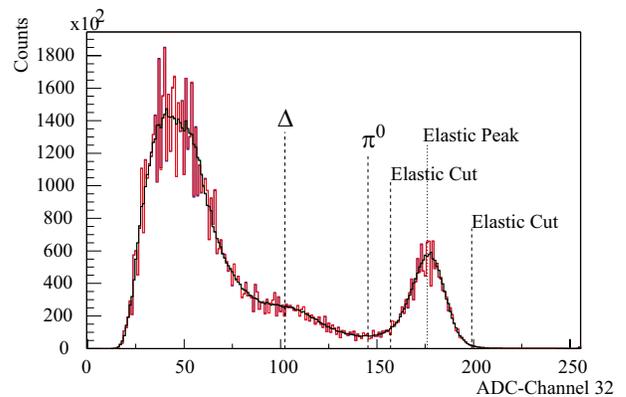}\\
      \caption{An energy spectrum of accepted particles from the
               hydrogen target as read directly from the hardware memory of the read out electronics
               of the lead fluoride calorimeter. For the solid black curve, the spectrum has been
               corrected for the differential nonlinearity of the ADC.
               The position of the elastic scattering peak,
               the threshold for $\pi^0$-production and the position of the $\Delta$-resonance
               is indicated as well as the lower and upper cut position for the extraction of
               $N_e^R$ and $N_e^L$ as described in the text.
                }\label{fig:a4spektrum}
    \end{center}
\end{figure}
calorimeter consisting of individual lead fluoride (PbF$_2$) crystals \cite{a4calori:maas:02}. The material
is a pure \cherenkov\ radiator and has been chosen for its fast timing characteristics and
its radiation hardness \cite{pbf2:achenbach:01}. This is the first time this material
is used in a large scale
calorimeter for a physics experiment. The crystals are dimensioned so that
an electron deposits $96\,\%$ of its total energy in an electro-magnetic shower
extending over a matrix of $3 \times 3$ crystals. Together with the readout electronics
this allows us a measurement of the particle energy with a resolution of $3.9\,\%/\sqrt{E}$
and a total dead time of 20~ns. At the time of the data taking, 511 channels of the
detector and the read out electronics were operational. The detector modules were located
in two sectors covering an azimuthal angle interval $\Delta \phi$ of 90$^{\circ}$
symmetrically around the beam axis.
The particle rate within the acceptance of this solid angle
is $\approx 50\times10^{6}$~s$^{-1}$. Due to the short dead time, the
losses due to double hits in the calorimeter are $\doublehitlosses\,\%$ at 20~$\mu$A.
%{\sc This low dead time is only possible because of the special read out electronics employed.}
The signals from each cluster of 9 crystals are summed and integrated
for 20~ns in an analogue summing and triggering circuit and digitized by a transient 8-bit
ADC. There is one summation, triggering, and digitization circuit
per crystal. The energy, helicity, and impact information are stored together in a three dimensional
histogram. Figure \ref{fig:a4spektrum} shows a typical
energy spectrum of scattered particles from the hydrogen target at an electron current of 20~$\mu$A .
It was taken during five minutes and is a direct output of the histogramming memory.
The elastic scattering peak is clearly isolated at the high end of the spectrum.\\
%
% Analysis of the data
%
The energy resolution $\sigma_E$ of each group of 9 crystals was determined
by analyzing such spectra. The number of elastic scattered electrons is
determined for each detector channel by integrating
the number of events in an interval from $\lowerenergycut\;\sigma_E$ above pion
production threshold to $\upperenergycut\;\sigma_E$ above the elastic peak
in each helicity histogram. The usage of those cuts ensures a clean separation between
elastic scattering and pion production or $\Delta$-excitation which has an unknown PV
cross section asymmetry. We determined the number of elastically scattered
electrons for each helicity state ($N_e^R$ and $N_e^L$) by summing over the inner
345 detector channels which are the centers of a full $3 \times 3$ crystal matrix.
The linearity of the PbF$_2$ detector system with respect to particle counting rates
and possible effects due to dead time were investigated by varying the beam current.
We calculate the raw normalized detector asymmetry as
$A_{\rm raw} = (N_e^R/\rho^R - N_e^L/\rho^L) / (N_e^R/\rho^R + N_e^L/\rho^L)$.
The possible dilution of the measured asymmetry by background
originating from the production of $\pi^0$ which
subsequently decays into two photons where one of the photons carries
almost the full energy of an elastic scattered electron was
estimated using Monte Carlo simulations to be much less than
$\gammafrompizerodilutionfactor\,\%$ and is neglected here.
The largest background comes from quasi elastic scattering at the thin aluminum entrance
and exit windows of the target cell. We have measured the aluminum quasielastic
event rate and calculated in static approximation a correction factor
for the aluminum of
\aluminumwindowcorectionfactor$\pm$ \aluminumwindowcorectionfactorerror \
giving a smaller value for the corrected asymmetry.\\
Corrections to false asymmetries arising from helicity correlated changes of beam parameters
were applied on a run by run base. The analysis was
based on the five min runs for which the counted elastic events in the PbF$_2$ detector were combined
with the correlated beam parameter and luminosity measurements. In the analysis we applied
reasonable cuts in order to exclude runs where the accelerator or parts of the
PbF$_2$ detector system were malfunctioning.
The analysis is based on a total of
$\totalnumberofhistograms \times 10^{6}$ histograms corresponding to
$\totalnumberofevents \times 10^{12}$ elastic scattering events.\\
%
% Extracting experimental result Aexp
%
We extracted an experimental asymmetry from
$A_{\rm exp}= A_{\rm raw} - a_1 A_I
                    - a_2 \Delta x
                    - a_3 \Delta y
                    - a_4 \Delta x'
                    -a_5 \Delta y'
                    - a_6 \Delta E_e$.
The six $a_i (i=1\ldots6)$ denote the correlation coefficients between the observed
false asymmetry and the electron current asymmetry $A_I$ ($a_1$), the
horizontal and vertical beam position differences $\Delta x$, $\Delta y$ ($a_2$, $a_3$),
the horizontal and vertical beam angle differences $\Delta x'$, $\Delta y'$ ($a_4$, $a_5$),
and the beam energy difference $\Delta E_e$ ($a_6$).
For the analysis, the correlation parameters $a_i$ were extracted by
multidimensional regression analysis from the data.
The $a_i$ have been calculated in addition from the geometry of the
precisely surveyed detector geometry and the $a_i$ calculated
using the two different methods agree very well within statistics.
%{\sc The direct calculation from geometry is possible
%for our setup since there is no imaging system between the target and
%the detector system. }
\\
%
% extracting Aphys
%
The experimental asymmetry was normalized to the electron beam polarization $P_e$
\begin{figure}%[htb]
    \begin{center}
      \includegraphics[width=0.45\textwidth]{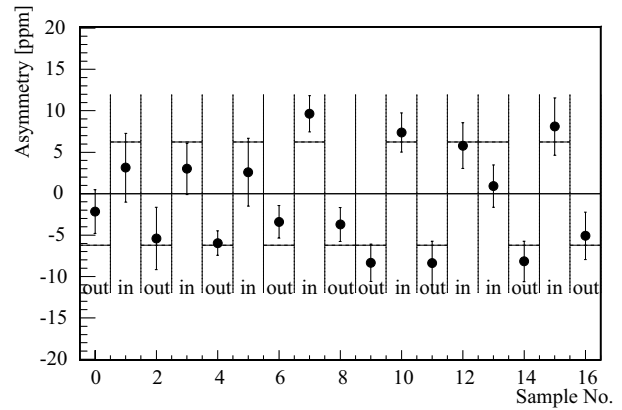}\\
      \caption{The extracted experimental asymmetries
      are shown with $\lambda /2$-plate in or out respectively as a function of
      the data sample. The dashed line represents the value of $A_0$ as described in the text.
      }\label{fig:a4results}
    \end{center}
\end{figure}
to extract the physics asymmetry, $A_{\rm phys}= A_{\rm exp}/ P_e$.
We have taken half of our data with a second $\lambda/2$-plate inserted
between the laser system and the GaAs crystal.
This reverses the polarization of the electron beam and allows a
stringent test of the understanding of systematic effects.
The effect of the plate can be seen in Fig.~\ref{fig:a4results}: the observed
asymmetry extracted from the different data samples changes sign,
which is a clear sign of parity violation if, as in our case, the target is
unpolarized.
Our measured result for the PV physics asymmetry in the scattering cross section of
polarized electrons on unpolarized protons at an average $Q^2$ value of $\qsquaredavaraged$~(GeV/c)$^2$
is    $A_{\rm phys} = (\experimentalasymmetryalucor \pm
                   \statisticalerroralucor \pm
                   \combinedsyspolerroralucor)$~ppm. The first error represents
the statistical accuracy, the second error the systematical uncertainties
including beam polarization.
The absolute accuracy of the experiment represents the most accurate measurement
of a PV asymmetry in
the elastic scattering of longitudinally polarized electrons on unpolarized
protons.
Table \ref{tab:results errors} gives an overview of the applied corrections and the
\begin{table}
\caption{Overview of the applied corrections and the sources of the experimental
error in the measured asymmetry.
\label{tab:results errors}}
\begin{ruledtabular}
\begin{tabular}{l r r}
                &\multicolumn{1}{c}{correction [ppm]}& \multicolumn{1}{c}{error [ppm]}   \\ \hline
Statistics      & \budgetstatisticcorr               & \budgetstatisticerr               \\ \hline
Target density, luminosity
                & \budgettargetdensLcorr             & \budgettargetdensLerr             \\
Target density, beam current
                & \budgettargetdensIcorr             & \budgettargetdensIerr             \\
Nonlinearity of LuMo
                & \budgetLUMOnonlincorr              & \budgetLUMOnonlinerr              \\
Dead time correction
                & \budgetquadcorr                    & \budgetquaderr                    \\
$A_I$           & \budgetasymIcorr                   & \budgetasymIerr                   \\
$\Delta E_e$    & \budgetenergycorr                  & \budgetenergyerr                  \\
$\Delta x$, $\Delta y$
                & \budgetposcorr                     & \budgetposerr                     \\
$\Delta x'$, $\Delta y'$
                & \budgetangcorr                     & \budgetangerr                     \\
Aluminum windows (H$_2$ target)
                & \aluminumwindowcorrection          & \aluminumwindowcorrectionerror    \\
Dilution from $\pi^0$ decay
                & \gammafrompizerodilutioncorrection & \gammafrompizerodilutioncorrectionerror     \\
$P_e$ measurement
                & \polarisationmeasurementcorrection & \polarisationmeasurementerror     \\
$P_e$ interpolation
                & \polarisationinterpolationcorrection & \polarisationinterpolationerror   \\ \hline
Systematic error
                &                                    & \combinedsyspolerroralucor
\end{tabular}
\end{ruledtabular}
\end{table}
contributions to the systematical error of the experimental asymmetry.\\
%
%  Interpretation
%
The interpretation of the measurement in terms of strangeness contribution is possible
by comparing the measured physics asymmetry $A_{\rm phys}$ with the averaged theoretical value
without strangeness contribution to the vector form factors $A_0$.
The difference $A_{\rm phys} - A_0$ is proportional to an averaged combination of the Sachs form factors
and the extracted value is
$G^s_E + \gesplusgmsaveragedfactor G^s_M = \gesplusgmsaveraged \pm \gesplusgmsaveragederror$.
If one uses the Dirac and Pauli form factors instead, the extracted value is
$F^s_1 + \fonesplusftwosaveragedfactor F^s_2 = \fonesplusftwosaveraged \pm \fonesplusftwosaveragederror$.
The solid line in Fig.~\ref{fig:gegmcombined} illustrates the possible combinations
of $G^s_E$ and $G^s_M$ given by our result on the
measured combination of $G^s_E + \gesplusgmsaveragedfactor G^s_M$ at
$Q^2=\qsquaredavaraged$~(GeV/c)$^2$. The hatched area represents the error on
$G^s_E + \gesplusgmsaveragedfactor G^s_M$.
The measured combination is small and is 1.2 standard deviations away from
zero, which clearly rules out the pole fit type of theoretical models
on the strangeness in the nucleon \cite{strangeness:jaffe:89,strangeness:hammer:96}.
From the published result on the measured asymmetry of the HAPPEX collaboration
\cite{happex:aniol:01} at $Q^2$=\qsquaredhappex~(GeV/c)$^2$ and $\theta_e = \happexangle^{\circ}$,
we recalculated the combination using our parametrization for the electro-magnetic form factors \cite{emformfactor:friedrich:03}
and yield
$G^s_E + \gesplusgmsfactorhappex G^s_M = \gesplusgmshappex \pm \gesplusgmserrorhappex$
and $F^s_1 + \fonesplusftwosfactorhappex F^s_2 = \fonesplusftwoshappex \pm \fonesplusftwoserrorhappex$.
%The similarity of the results from HAPPEX and A4 (this work)
%leads us to a combination of the two results with the aim of gaining
%significance in an estimate on $F^s_1$ and $G_E^s$ respectively.
%{ \sc The theoretical estimates of a strangeness contribution evaluated
%in different models including lattice calculation show despite all differences
%in sign and absolute size the common feature, that the $Q^2$ evolution of the
%strangeness contribution to the vector form factors of the nucleon is rather
%smooth \cite{strangereview:beck:01}.}
In lack of more detailed information
we make the ad hoc assumption, that $F^s_2$ in the $Q^2$-range between 0.1~(GeV/c)$^2$
and 0.5~(GeV/c)$^2$ can be approximated by $F^s_2 = \ftwosassumption \pm \ftwosassumptionerror$,
corresponding to $G^s_M = \gmsassumption \pm \gmsassumptionerror$.
This assumption is guided by the fact that this value covers within two standard deviations
all theoretical estimates as well as the SAMPLE result from \cite{sample:hasty:00}.
This yields a value for $G_E^s$ evaluated from our measurement of
$G_E^s(Q^2 = \qsquaredavaraged$~(GeV/c)$^2) = \finalgesafour \pm \finalgesafourerror$ and
$F_1^s(Q^2 = \qsquaredavaraged$~(GeV/c)$^2) = \finalfonesafour \pm \finalfonesafourerror$.
If one makes the further approximation of
neglecting the $Q^2$-dependence in $G_E^s$, we can combine
our result with the recalculated HAPPEX result by calculating the weighted average and yield an estimate of
$F_1^s(0.1$~(GeV/c)$^2 < Q^2 < 0.5$~(GeV/c)$^2) = \finalfonescombined \pm \finalfonescombinederror$ and
$G_E^s(0.1$~(GeV/c)$^2 < Q^2 < 0.5$~(GeV/c)$^2) = \finalgescombined \pm \finalgescombinederror$.\\
The significance level of \finalfonessigma~$\sigma$ for $F^s_1$ and \finalgessigma~$\sigma$ for $G_E^s$
which we reach using the assumption described above, leads us
to the conclusion, that the combination of our measurements presented here with the earlier work
of the HAPPEX collaboration shows for the first time evidence for the observation of
a contribution of the strange quarks to the electric vector form factor of the nucleon.\\
%
%  No Outlook
%
%
%Acknowledgments
%
This work is supported by the Deutsche Forschungsgemeinschaft
in the framework of the SFB 201, SPP 1034, by the IN2P3 of CNRS
and in part by the US Department of Energy.
We are indebted to K.H. Kaiser and the whole MAMI crew for their
tireless effort to provide us with good electron beam. We also
would like to thank the A1 Collaboration for the use of the
\moeller\ polarimeter.

\bibliography{prl_855MeV_long}

\end{document}